\documentclass[pra,twocolumn,tbtags,superscriptaddress,floatfix]{revtex4}
\usepackage{amsmath}
\usepackage{graphicx}
\usepackage{amssymb}
\usepackage{color,soul}
\usepackage[T1]{fontenc}
\usepackage{ae,aecompl}
\usepackage{natbib}

\begin{document}

\title{Thermometry and memcapacitance with qubit-resonator system}

\begin{abstract}
We study theoretically dynamics of a driven-dissipative qubit-resonator
system. Specifically, a transmon qubit is coupled to a transmission-line
resonator; this system is considered to be probed via a resonator, by means
of either continuous or pulsed measurements. Analytical results obtained in
the semiclassical approximation are compared with calculations in the
semi-quantum theory as well as with the previous experiments. We demonstrate
that the temperature dependence of the resonator frequency shift can be used
for the system thermometry and that the dynamics, displaying
pinched-hysteretic curve, can be useful for realization of memory devices,
the quantum memcapacitors.
\end{abstract}

\date{\today }
\author{S.~N.~Shevchenko}
\affiliation{B.~I.~Verkin Institute for Low Temperature Physics and Engineering, Kharkov,
Ukraine}
\affiliation{V.~N.~Karazin Kharkov National University, Kharkov, Ukraine}
\author{D.~S.~Karpov}
\affiliation{B.~I.~Verkin Institute for Low Temperature Physics and Engineering, Kharkov,
Ukraine}
\maketitle

\section{Introduction}

\ \ The key object of the up-to-date circuit QED is the system comprised of
a qubit coupled to the quantum transmission-line resonator \cite{Koch07}.
Such systems are useful for both studying fundamental quantum phenomena and
for quantum information protocols including control, readout, and memory
\cite{Ashhab10, Wendin16}. Realistic QED system includes also electronics
for driving and probing, while the general consideration should include in
addition the inevitable dissipative environment and non-zero temperature.

In many cases, the temperature can be assumed equal to zero. However, there
are situations when it is important both to take into account and to monitor
the effective temperature \cite{Giazotto06, Albash17}. One of the reasons is
that it is a variable value, which depends on several factors \cite%
{Wilson10, Forn-Diaz16, Stehlik16}, for example it significantly varies with
increasing driving power. Different aspects of the thermometry involving
qubits were studied in Refs.~[%
\onlinecite{Palacios-Laloy09, Fink10, Brunelli11, Higgins13, Ashhab14,
Jevtic15, Ahmed18}].

Even though our consideration is quite general and can be applied to other
types of qubit-resonator systems, including semiconductor qubits \cite{Mi17}%
, for concreteness we concentrate on a transmon-type qubit in a cavity, of
which the versatile study was presented in Ref.~[\onlinecite{Bianchetti09}].
These systems were studied for different perspectives, recently including
such an elaborated phenomena as the Landau-Zener-St\"{u}ckelberg-Majorana
interference \cite{Gong16}. The impact of the temperature was studied in
Ref.~[\onlinecite{Fink10}], however the authors were mainly interested in
the resonator temperature. Here we explicitly take into account the non-zero
effective temperature impact on both resonator and qubit. First, we obtain
simplified but transparent analytical expressions for the transmission
coefficient in the semi-classical approximation, which ignores the
qubit-resonator correlations. Such semiclassical approach is useful, but its
validity should be checked \cite{Remizov16}. For this reason, we further
develop our calculations, by taking into account the qubit-resonator
correlators.

Having obtained agreement with previous experiments, such as the ones in
Refs.~[\onlinecite{Bianchetti09, Jin15, Pietikainen17a}], we also consider
another emergent application, for memory devices. Different types of memory
devices, such as memcapacitors and meminductors, were introduced in addition
to memristors \cite{Diventra09, Pershin11}. See also Refs.~[%
\onlinecite{Peotta14,
Guarcello16, Guarcello17}] for different proposals of superconducting memory
elements. Quantum versions of memristors, memcapacitors, and meminductors
were discussed in Refs.~[%
\onlinecite{Pfeiffer16, Shevchenko16, Salmilehto16,
Li16, Sanz17}]. In particular, in Ref.~[\onlinecite{Shevchenko16}] it was
suggested that a charge qubit can behave as a quantum memcapacitor. We
consider here a transmon qubit in a cavity, instead of a charge qubit, as a
possible candidature for the realization of the quantum memcapacitor. For
this, we demonstrate that the transmon-resonator system can be described by
the relations defining a memcapacitor.

Overall, the paper is organized as following. In Sec.~II we consider the
driven qubit-resonator system probed via quadratures of the transmitted
field. This is developed by taking temperature into account in Sec.~III,
where continuous measurements are considered. While we compare our results
with Ref.~[\onlinecite{Bianchetti09}], our approach there (also presented in
Appendix~A), was the semiclassical theory, valid for both dispersive and
resonant cases. Importantly, we verify the results with the more elaborated
calculations, taking into account two-operator qubit-resonator correlators,
of which the details are presented in Appendix~B. Section~IV is devoted to
the case of single-shot pulsed measurements. In Sec.~V, we consider cyclic
dynamics with hysteretic dependencies, needed for emergent memory
applications.

\section{Time-dependence of the quadratures}

The qubit-resonator system we consider in the circuit-QED realization, as
studied in Refs.~[\onlinecite{Koch07, Bianchetti09}]. The qubit is the
transmon formed by an effective Josephson junction and the shunt capacitance
$C_{\mathrm{B}}$; it is capacitively coupled to the transmission-line
resonator via $C_{\mathrm{g}}$, as shown in the inset in Fig.~\ref%
{Fig:vs_time_1}. The resonator is driven via $C_{\mathrm{in}}$ and measured
value is the transmitted electromagnetic field after $C_{\mathrm{out}}$. In
addition, the effective Josephson junction stands for the loop with two
junctions controlled by an external magnetic flux $\Phi $; the respective
Josephson capacitance and energy are denoted in the scheme with $C_{\mathrm{J%
}}$\ and $E_{\mathrm{J}}$. The qubit characteristic charging energy is $E_{%
\mathrm{c}}=e^{2}/2C_{\Sigma }$ with $C_{\Sigma }=C_{\mathrm{J}}+C_{\mathrm{B%
}}+C_{\mathrm{g}}$.

\begin{figure}[t]
\includegraphics[width=8cm]{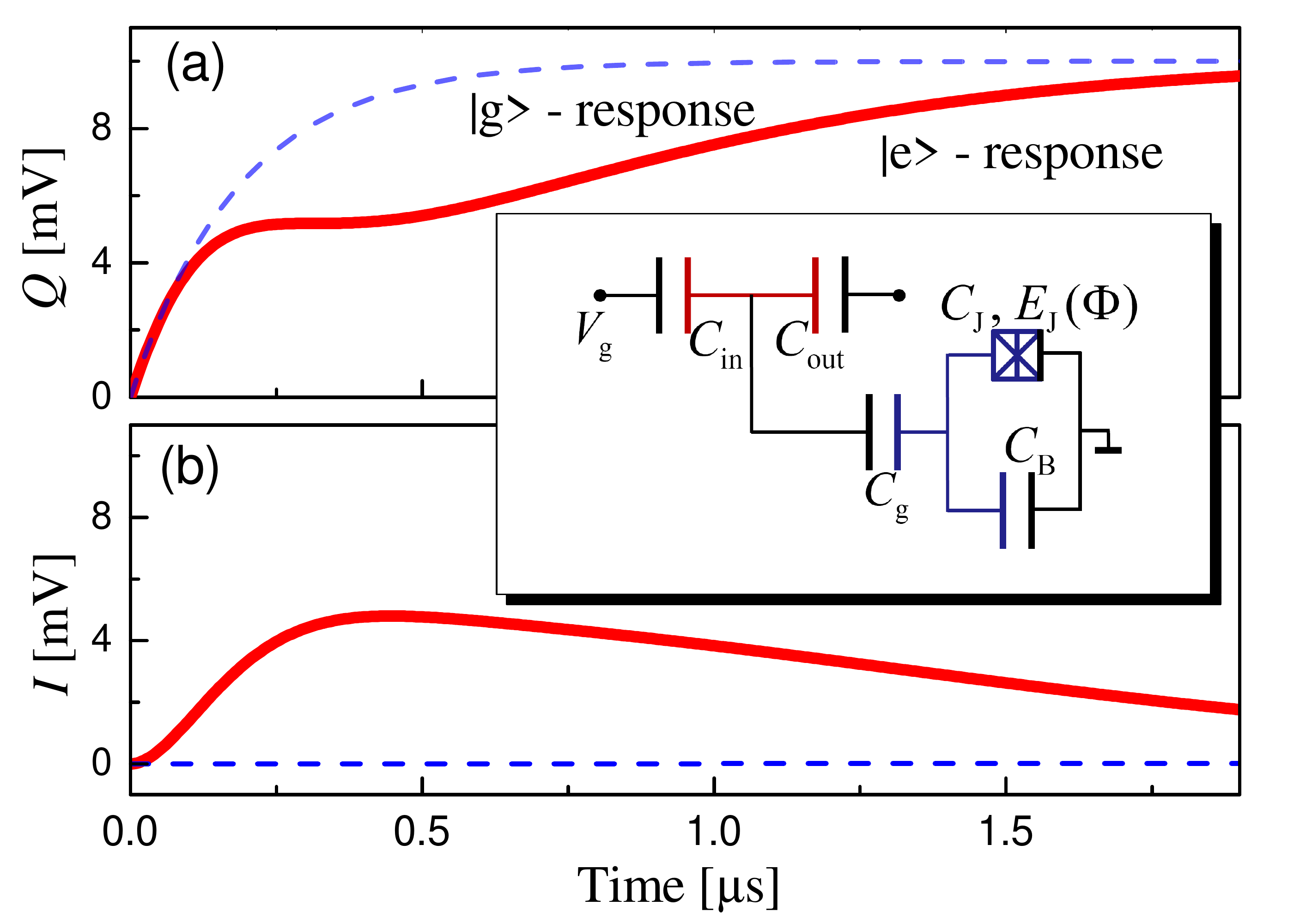}
\caption{Time evolution of the quadratures, $Q$ (a) and $I$ (b), for the
parameters of Ref.~[\onlinecite{Bianchetti09}] for the two situations, when
the qubit was initialized in either the ground or excited state, denoted as
\textquotedblleft $\left\vert g\right\rangle $-response\textquotedblright\
and \textquotedblleft $\left\vert e\right\rangle $-response%
\textquotedblright , respectively. The inset presents the scheme of the
transmon-type qubit coupled to the transmission-line resonator.}
\label{Fig:vs_time_1}
\end{figure}

The driven transmon-resonator system \cite{Koch07, Bianchetti09, Bishop09}
is described by the Jaynes-Cummings Hamiltonian \cite{Schleich}
\begin{eqnarray}
H &=&\hbar \omega _{\mathrm{r}}a^{\dag }a+\hbar \frac{\omega _{\mathrm{q}}}{2%
}\sigma _{z}+\hbar \mathrm{g}\left( \sigma a^{\dag }+\sigma ^{\dag }a\right)
+  \label{JC} \\
&&+\hbar \xi \left( a^{\dag }e^{-i\omega t}+ae^{i\omega t}\right) .  \notag
\end{eqnarray}%
Here the transmon is considered in the two-level approximation, described by
the energy distance $\hbar \omega _{\mathrm{q}}$ between the levels and the
Pauli operators $\sigma _{i}$ and $\sigma _{\pm }=\left( \sigma _{x}\pm
i\sigma _{y}\right) /2$, where we rather use the ladder-operator notations $%
\sigma \equiv \sigma _{-}$ and $\sigma ^{\dag }\equiv \sigma _{+}$; the
resonator is described by the resonant frequency $\omega _{\mathrm{r}}$ and
the annihilation operator $a$; the transmon-resonator coupling constant $%
\mathrm{g}$ relates to the bare coupling $\mathrm{g}_{0}$ as $\mathrm{g}=%
\mathrm{g}_{0}\sqrt{E_{\mathrm{c}}/\left\vert \Delta -E_{\mathrm{c}%
}\right\vert }$ with $\Delta =\hbar \left( \omega _{\mathrm{q}}-\omega _{%
\mathrm{r}}\right) $ (this renormalization is due to the virtual transitions
through the upper transmon's states); the probing signal is described by the
amplitude $\xi $ and frequency $\omega $.

The system's dynamics obeys the master equation, which is described in
Appendix A. There, it is demonstrated that the Lindblad equation for the
density matrix can be rewritten\ as an infinite set of equations for the
expectation values. In Refs.~[\onlinecite{Bianchetti09, Shevchenko14}] the
set of equations was reduced to six complex equations for the single
expectation values and the two-operator correlators. Meanwhile, many
quantum-optical phenomena can be described within the semiclassical theory,
assuming all the correlation functions to factorize (e.g. Refs.~[%
\onlinecite{Mu92, Hauss08, Andre09, Macha14}]). This approach results in
that the system's dynamics is described by the set of three equations only,
Eqs.~(\ref{Maxwell-Bloch}), which are more suitable for analytic
consideration, as we will see below. This was also analyzed in Ref.~[%
\onlinecite{Shevchenko14}]; in particular, the robustness of the
semiclassical approximation was demonstrated even in the limit of small
photon number, at small probing amplitude $\xi $.

The observable value can be either transmission signal amplitude or the
quadrature amplitudes. The quadratures of the transmitted field $I$ and $Q$
are related to the cavity field $\left\langle a\right\rangle $ as following
\cite{Bianchetti09, Bishop09}%
\begin{equation}
I=2V_{0}\,\,\text{Re}\left\langle a\right\rangle \text{, \ }Q=2V_{0}\,\,%
\text{Im}\left\langle a\right\rangle ,  \label{IQ}
\end{equation}%
where $V_{0}$ is a voltage related to the gain of the experimental
amplification chain \cite{Bishop09} and it is defined as \cite{Bianchetti09}
$V_{0}^{2}=Z\hbar \omega _{\mathrm{r}}\varkappa /4$ with $Z$ standing for
the transmission-line impedance. The transmission amplitude $A$ is given
\cite{Bishop09, Macha14} by the absolute value of $\left\langle
a\right\rangle $
\begin{equation}
A=\sqrt{I^{2}+Q^{2}}=2V_{0}\left\vert \left\langle a\right\rangle
\right\vert .  \label{A}
\end{equation}

\begin{figure}[t]
\includegraphics[width=8cm]{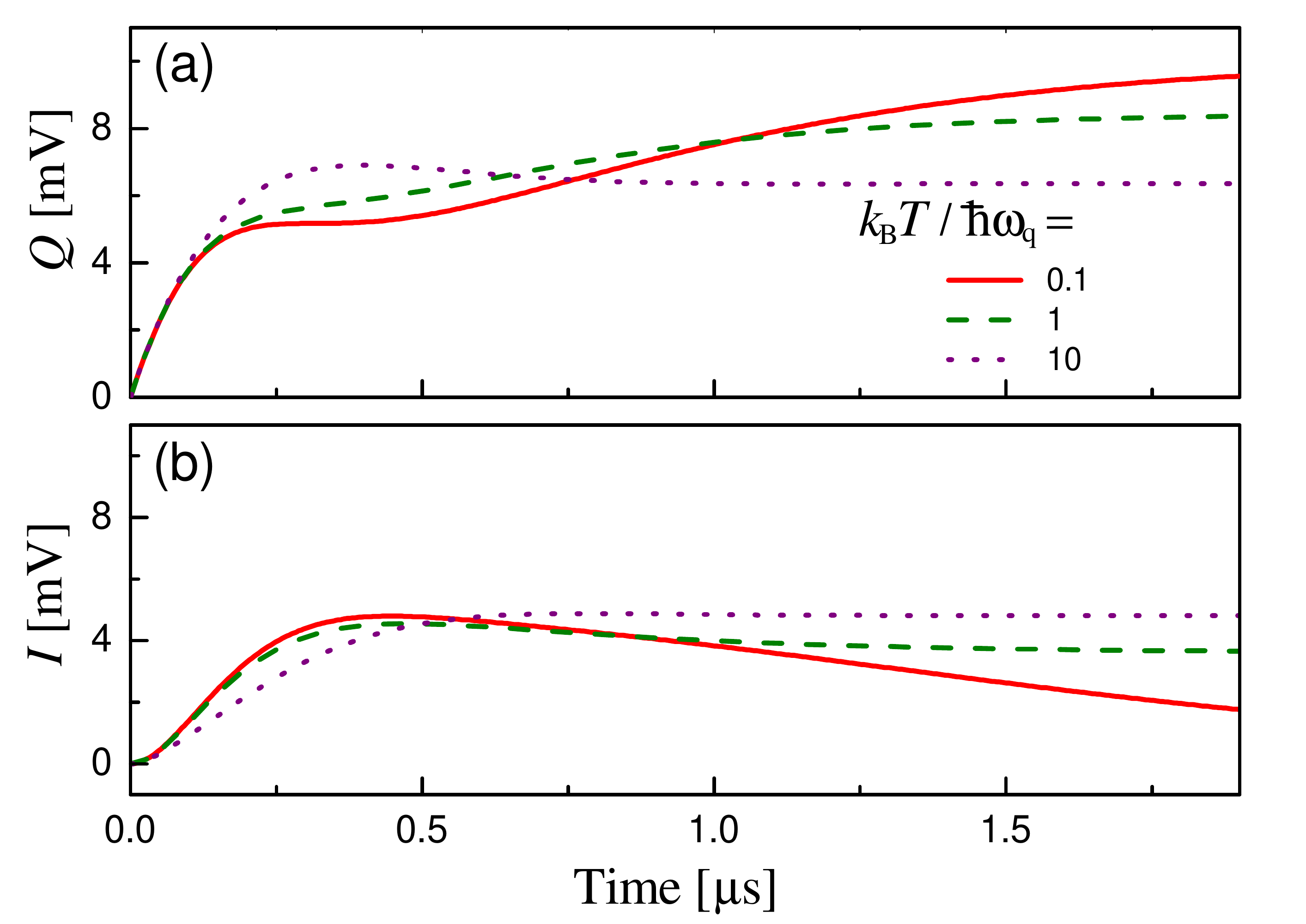}
\caption{Time evolution of the quadratures, $Q$ (a) and $I$ (b), for
non-zero temperature $T$. The situation when the qubit was initialized in
the excited state is considered. The parameters are the same as in Fig.~%
\protect\ref{Fig:vs_time_1}, besides the temperature, so that the solid red
curves for the low temperature repeat the ones from the previous figure.}
\label{Fig:vs_time_2}
\end{figure}

As an illustration of the semiclassical theory, presented in more detail in
Appendix A, consider the experimental realization in Ref.~[%
\onlinecite{Bianchetti09}]. There, the qubit was initialized in either
ground or excited state and then, by means of either continuous or pulsed
measurements, the quadratures of the transmitted field were probed.
Correspondingly, we make use of Eqs.~(\ref{IQ})~and~(\ref{Maxwell-Bloch}),
which include the resonator relaxation rate $\varkappa $\ and the qubit
decoherence rate $\Gamma _{2}=\Gamma _{\phi }+\Gamma _{1}/2$ with $\Gamma
_{\phi }$ and $\Gamma _{1}$ being the intrinsic qubit pure dephasing and
relaxation rates. We take the following parameters \cite{Bianchetti09}: $%
\omega _{\mathrm{r}}/2\pi =6.4425$~GHz, $\omega _{\mathrm{q}}/2\pi =4.01$%
~GHz, $\mathrm{g}_{0}/2\pi =134$~MHz,$\ \varkappa /2\pi =1.7$~MHz, $\Gamma
_{1}/2\pi =0.2$~MHz, $\Gamma _{2}=\Gamma _{1}/2$, $E_{\mathrm{c}}/h=232$%
~MHz, and $V_{0}=5$~mV, where the latter was chosen as a fitting parameter.
The results for low temperature (i.e. for $k_{\mathrm{B}}T\ll \hbar \omega
_{_{\mathrm{q}}}$) are presented in Fig.~\ref{Fig:vs_time_1}. Note the
agreement with the experimental observations in Ref.~[%
\onlinecite{Bianchetti09}]; see also detailed calculations in Appendix~B
below. There, in Ref.~[\onlinecite{Bianchetti09}] it is discussed in detail
that the relaxation of the quadratures is determined for the ground-state
formulation by the resonator rate $\varkappa $ only, while for the
excited-state formulation this is determined by the collaborative evolution
of the qubit-resonator system. For example, one can observe that the
relaxation of the quadratures in Fig.~\ref{Fig:vs_time_1} for the
\textquotedblleft $\left\vert e\right\rangle $-response\textquotedblright\
happens in two stages, during the times $T_{\varkappa }=2\pi /\varkappa
\simeq 0.6\mu $s and $T_{1}=2\pi /\Gamma _{1}\gg T_{\varkappa }$.

\section{Thermometry with continuous measurements}

In previous Section we calculated the low-temperature behaviour of the
observable quadratures for the qubit-resonator system and illustrated this
in Fig.~\ref{Fig:vs_time_1}. Having obtained the agreement with the
experimental observations of Ref.~\cite{Bianchetti09}, we can proceed with
posing other problems for the system. Consider now the sensitivity of the
system to the changes of temperature. How the behaviour of the observables
changes? Is this useful for a single-qubit thermometry?\ To respond to such
questions, we describe below both dynamical and stationary behaviour for
non-zero temperature.

In Fig.~\ref{Fig:vs_time_2} we plot the time evolution of the quadratures
for the same parameters as in Fig.~\ref{Fig:vs_time_1} besides the
temperature, which now is considered non-zero. Figure~\ref{Fig:vs_time_2}
demonstrates that both evolution and stationary values (at long times,
independent of initial conditions) are strongly temperature dependent.

To further explore the temperature dependence, we now consider the
steady-state measurements. In equilibrium, the observables are described by
the steady-state values of $\left\langle a\right\rangle $, $\left\langle
\sigma \right\rangle $, and $\left\langle \sigma _{z}\right\rangle $. The
steady-state solution for the weak driving amplitude in the semiclassical
approximation is the following (for details see Appendix A):

\begin{equation}
\left\langle a\right\rangle =-\xi \frac{\delta \omega _{\mathrm{q}}^{\prime }%
}{\left\langle \sigma _{z}\right\rangle \mathrm{g}^{2}+\delta \omega _{%
\mathrm{q}}^{\prime }\delta \omega _{\mathrm{r}}^{\prime }},  \label{a}
\end{equation}%
where
\begin{eqnarray}
\delta \omega _{\mathrm{r}}^{\prime } &=&\omega _{\mathrm{r}}-\omega -i\frac{%
\varkappa }{2},\text{ }\delta \omega _{\mathrm{q}}^{\prime }=\omega _{%
\mathrm{q}}-\omega -i\frac{\Gamma _{2}}{z_{0}}, \\
z_{0} &=&\tanh \left( \frac{\hbar \omega _{_{\mathrm{q}}}}{2k_{\mathrm{B}}T}%
\right) .  \notag
\end{eqnarray}%
In equilibrium, the qubit energy-level populations are defined by the
temperature $T$: $\left\langle \sigma _{z}\right\rangle =-z_{0}$ \cite{Jin15}%
. Importantly, formula~(\ref{a}) bears the information about the qubit
temperature and via formula (\ref{A}) brings this dependence to the
observables.

Formula~(\ref{a}) is quite general. To start with, for an isolated resonator
(without qubit) at $\mathrm{g}=0$ this gives
\begin{equation}
\left\vert \left\langle a\right\rangle \right\vert ^{2}=\xi ^{2}\frac{1}{%
\delta \omega _{\mathrm{r}}^{2}+\varkappa ^{2}/4},  \label{a0}
\end{equation}%
which defines the resonator width.

Consider now the \textit{dispersive} limit, where $\Delta /\hbar \equiv
\omega _{\mathrm{q}}(\Phi )-\omega _{\mathrm{r}}\gg \mathrm{g}/h,\,\,\delta
\omega _{\mathrm{r}}$. Then we have for the transmission amplitude
\begin{equation}
\left\vert \left\langle a\right\rangle \right\vert ^{2}\approx \xi ^{2}\frac{%
\Delta ^{2}}{\left( \left\langle \sigma _{z}\right\rangle \mathrm{g}%
^{2}+\Delta \delta \omega _{\mathrm{r}}\right) ^{2}+\Delta ^{2}\varkappa
^{2}/4}.  \label{general}
\end{equation}%
This, in particular, gives the maxima for the transmission at%
\begin{equation}
\delta \omega _{\mathrm{r}}=-\left\langle \sigma _{z}\right\rangle \frac{%
\mathrm{g}^{2}}{\Delta }\equiv -\left\langle \sigma _{z}\right\rangle \chi .
\label{dispersive}
\end{equation}%
Then, for the ground/excited states with $\left\langle \sigma
_{z}\right\rangle =\mp 1$, one obtains the two dispersive shifts for the
maximal transmission, $\delta \omega _{\mathrm{r}}=\pm \chi =\pm \mathrm{g}%
_{0}^{2}E_{\mathrm{c}}/\Delta (\Delta -E_{\mathrm{c}})$, respectively. In
thermal equilibrium, equation~(\ref{dispersive}) for the resonance frequency
shift gives $\delta \omega _{\mathrm{r}}(T)=\frac{\mathrm{g}^{2}}{\Delta }%
\tanh \left( \frac{\hbar \omega _{_{\mathrm{q}}}}{2k_{\mathrm{B}}T}\right) $.

\begin{figure}[t]
\includegraphics[width=8cm]{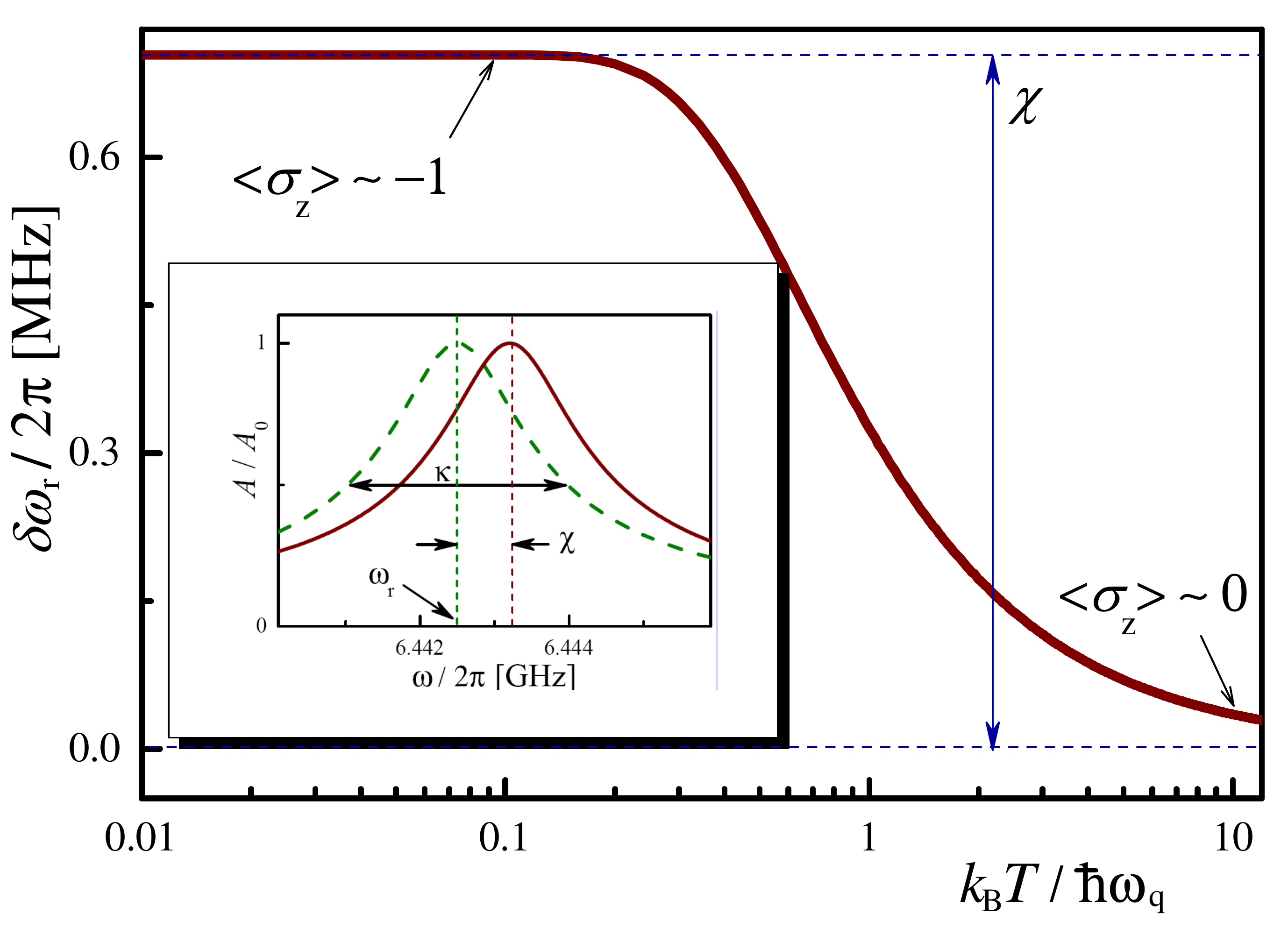}
\caption{Transmission amplitude and the frequency shift. First, the inset
shows the transmission amplitude $A$ versus the frequency $\protect\omega $
when the qubit is either in the ground state (solid line) or in the excited
state (dashed line). Then, the main panel demonstrates the frequency shift $%
\protect\delta \protect\omega _{\mathrm{r}}=\protect\omega _{\mathrm{r}}-%
\protect\omega $, corresponding to the frequency $\protect\omega $ at which
the transmission is maximal, plotted as a function of temperature $T$. Here
the transmission amplitude is normalized with $A_{0}=4V_{0}\mathrm{g}\protect%
\xi /\varkappa $. }
\label{Fig:phase_shift}
\end{figure}

Making use of Eqs.~(\ref{A})~and~(\ref{general}) in thermal equilibrium,
when $\left\langle \sigma _{z}\right\rangle =-z_{0}$, in the inset in Fig.~%
\ref{Fig:phase_shift} we plot the frequency dependence of the transmission
amplitude for the parameters of Ref.~[\onlinecite{Bianchetti09}]. We plot
two cuves, where the solid one corresponds to a low-temperature limit ($%
z_{0}=1$) with the system in the ground state, while the dashed line is
plotted in a high-temperature limit ($z_{0}=0$), when the system is in the
superposition of the ground and excited state. The maximal frequency shift
is denoted with $\chi $. Note that the low-temperature limit (solid line in
the inset), with $\left\langle \sigma _{z}\right\rangle \sim -1$,
corresponds to the ground state, while the high-temperature limit (dashed
line), with $\left\langle \sigma _{z}\right\rangle \sim 0$, is equivalent to
the absence of the qubit, at $\mathrm{g}=0$.

For varying temperature, the frequency shift is plotted in the main panel of
Fig.~\ref{Fig:phase_shift}, for the parameters of Ref.~[%
\onlinecite{Bianchetti09}]. We note that similar dependence can be found in
Fig.~4.2 of Ref.~[\onlinecite{Bianchetti10}]; the difference is in that in
the case of Refs.~[\onlinecite{Bianchetti09, Bianchetti10}] similar change
of $\left\langle \sigma _{z}\right\rangle $ from $-1$ to $0$ was due to
varying the driving power. When driven with low power, qubit stayed in the
ground state with $\left\langle \sigma _{z}\right\rangle =-1$, while with
increasing the power its stationary state tended to equally populated states
with $\left\langle \sigma _{z}\right\rangle =0$. Also, to this case of
varying the qubit driving, we further devote Appendix~C.

The temperature dependence in Fig.~\ref{Fig:phase_shift} becomes apparent at
$T\geq T^{\ast }$, where $T^{\ast }=0.1\hbar \omega _{_{\mathrm{q}}}/k_{%
\mathrm{B}}$ is the characteristic temperature, which, say, for $\omega _{_{%
\mathrm{q}}}/2\pi =4$~GHz is quite low, $T^{\ast }=20$ mK. This means that
such measurements may be useful for realizing the \textit{one-qubit
thermometry} for $T\geq T^{\ast }$.

\begin{figure}[t]
\includegraphics[width=8cm]{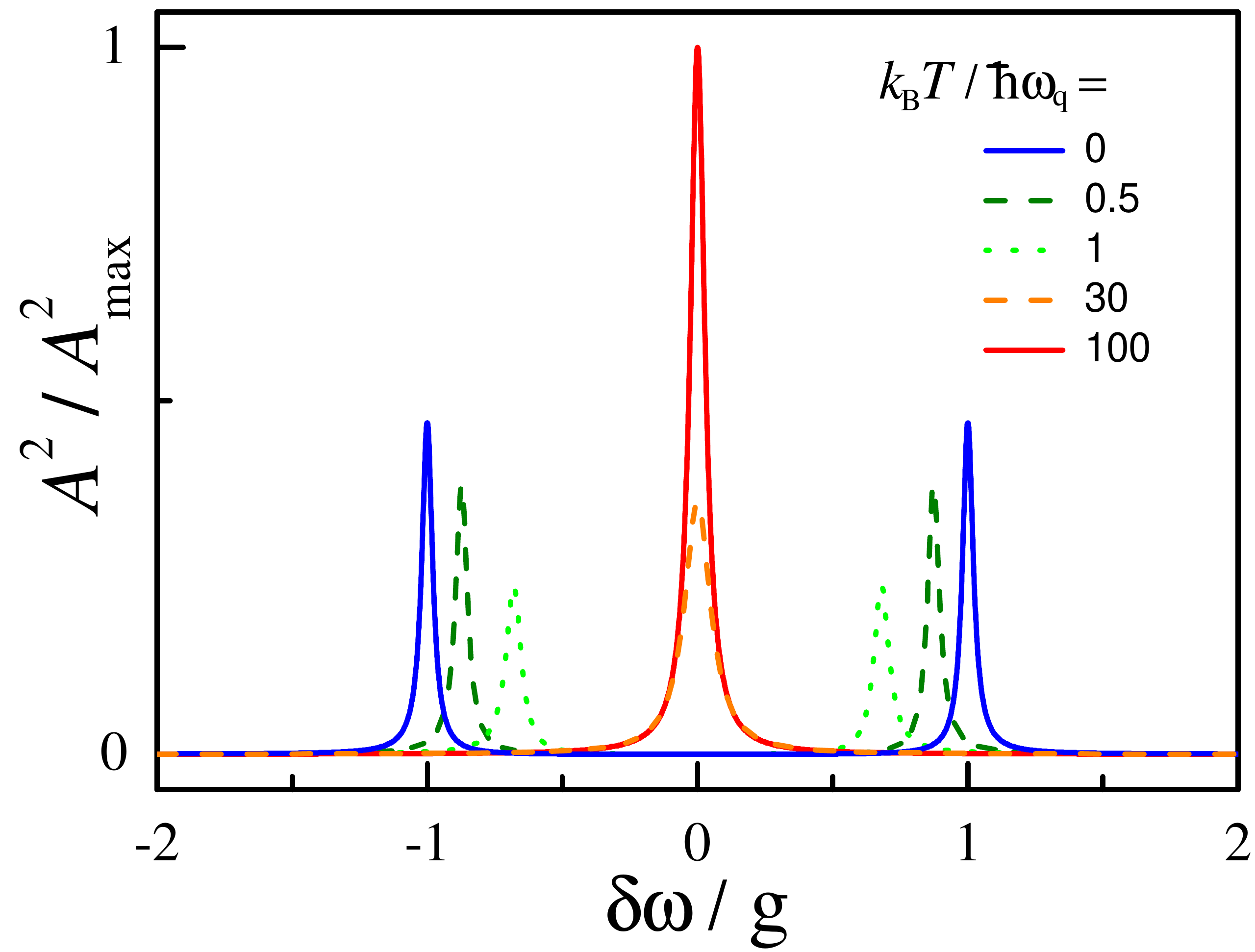}
\caption{Transmission $A^{2}$, normalized with its maximal value $A_{\max }$%
, versus the frequency shift $\protect\delta \protect\omega $ for different
values of temperature $T$.}
\label{Fig:resonant}
\end{figure}

It is important to note that Eq.~(\ref{a}) was obtained without making use
of the dispersive limit, and thus this is applicable to the opposite limit.
Consider in this way $\omega _{_{\mathrm{q}}}(\Phi )=\omega _{_{\mathrm{r}}}$%
, which is the \textit{resonant} limit, $\Delta =0$. With equal detunings
for both qubit and resonator, $\omega _{\mathrm{q}}-\omega =\omega _{\mathrm{%
r}}-\omega \equiv \delta \omega $, we can use the formula for the photon
operator, Eq.~(\ref{a}), which gives
\begin{equation}
\left\vert \left\langle a\right\rangle \right\vert ^{2}\approx \xi ^{2}\frac{%
\delta \omega ^{2}+\Gamma _{2}^{2}/z_{0}^{2}}{\left( \left\langle \sigma
_{z}\right\rangle \mathrm{g}^{2}+\delta \omega ^{2}\right) ^{2}+\delta
\omega ^{2}\left( \Gamma _{2}/z_{0}+\varkappa /2\right) ^{2}}.
\label{resonant_limit}
\end{equation}%
With this we plot the transmission amplitude as a function of the frequency
detuning in Fig.~\ref{Fig:resonant} for different temperatures. Formula~(\ref%
{resonant_limit}) describes maxima, which, assuming large cooperativity $%
\mathrm{g}^{2}/\Gamma _{2}\varkappa \gg 1$, are situated at $\delta \omega
=0 $ (the high-temperature peak) and at
\begin{equation}
\delta \omega \approx \pm \mathrm{g}\tanh ^{1/2}\left( \frac{\hbar \omega
_{_{\mathrm{q}}}}{2k_{\mathrm{B}}T}\right) .  \label{dw}
\end{equation}%
The latter formula, in particular, in the low-temperature limit describes
the peaks at $\delta \omega =\pm \mathrm{g}$, which is known as the vacuum
Rabi splitting. Note that recently such vacuum Rabi splitting was also
demonstrated in silicon qubits in Ref.~[\onlinecite{Mi17}]. With increasing
the qubit temperature, equation (\ref{dw}) means the temperature-dependent
resonance-frequency shift. We note, that this shift is again described by
the factor $\left\langle \sigma _{z}\right\rangle =-z_{0}$. If we assume
here the qubit in the ground state, $\left\langle \sigma _{z}\right\rangle
=-1$, then the increase of the temperature would result in suppressing the
peaks at $\delta \omega =\pm \mathrm{g}$, without their shift, in agreement
with Ref.~[\onlinecite{Fink10}].

\section{Thermometry with pulsed measurements}

Above we have considered the case when the measurement is done in a weak
continuous manner. Then, the resonator probes the averaged qubit state,
defined by $\left\langle \sigma _{z}\right\rangle $, and changing the qubit
state resulted in shifting the position of the resonant transmission.
Alternatively, the measurements can be done with the single-shot readout
\cite{Reed10, Jin15, Jerger16b, Jerger16, Reagor16}. In this case, in each
measurement, the resonator would see the qubit in either the ground or
excited state, with $\left\langle \sigma _{z}\right\rangle $ equal to $-1$
or $1$, respectively \cite{Vijay10}. Probability of finding the qubit in the
excited state is $P_{+}$ and in the ground state: $P_{-}=1-P_{+}$. Then, the
weighted (averaged over many measurements) transmission amplitude can be
calculated as following%
\begin{equation}
A=P_{-}A_{-}+P_{+}A_{+},
\end{equation}%
where $A_{\pm }$ describe the transmission amplitudes calculated for $%
\left\langle \sigma _{z}\right\rangle =\pm 1$, respectively, as given by
Eq.~(\ref{general}).

We may now consider two cases, of a qubit driven resonantly and when the
excitation happens due to the temperature. In the former case, when a qubit
is driven with frequency $\omega _{\mathrm{d}}=\omega _{\mathrm{q}}$ and
amplitude $\hbar \Omega $, the excited qubit state is populated with the
probability
\begin{eqnarray}
P_{+}(\Omega ) &=&\frac{1}{2}\left[ 1+\overline{\Omega }^{-2}\right] ^{-1},
\label{resonant} \\
\overline{\Omega } &=&\frac{1}{2}\hbar \Omega \sqrt{T_{1}T_{2}}.  \notag
\end{eqnarray}%
This is obtained from the full formula for a qubit excited near the resonant
frequency \cite{Shevchenko14}:%
\begin{equation}
P_{+}=\frac{1}{2}\frac{\omega _{\mathrm{q}}^{2}J_{1}^{2}\left( \frac{\Omega
}{\omega _{\mathrm{d}}}\right) }{\omega _{\mathrm{q}}^{2}J_{1}^{2}\left(
\frac{\Omega }{\omega _{\mathrm{d}}}\right) +\frac{T_{2}}{T_{1}}\left(
\omega _{\mathrm{q}}-\omega _{\mathrm{d}}\right) ^{2}+\frac{1}{T_{1}T_{2}}},
\label{non-resonant}
\end{equation}%
where we then take $\omega _{\mathrm{d}}=\omega _{\mathrm{q}}$ and $%
J_{1}(x)\approx x/2$.

In thermal equilibrium the upper-level occupation probability is defined by
the Maxwell-Boltzmann distribution, $\left\langle \sigma _{z}\right\rangle
=-z_{0}$ \cite{Jin15}, so that $P_{+}=\frac{1}{2}\left[ 1+\left\langle
\sigma _{z}\right\rangle \right] $\ or%
\begin{equation}
P_{+}(T)=\frac{1}{2}\left[ 1-\tanh \left( \frac{\hbar \omega _{_{\mathrm{q}}}%
}{2k_{\mathrm{B}}T}\right) \right] .  \label{T-excited}
\end{equation}

\begin{figure}[t]
\includegraphics[width=8cm]{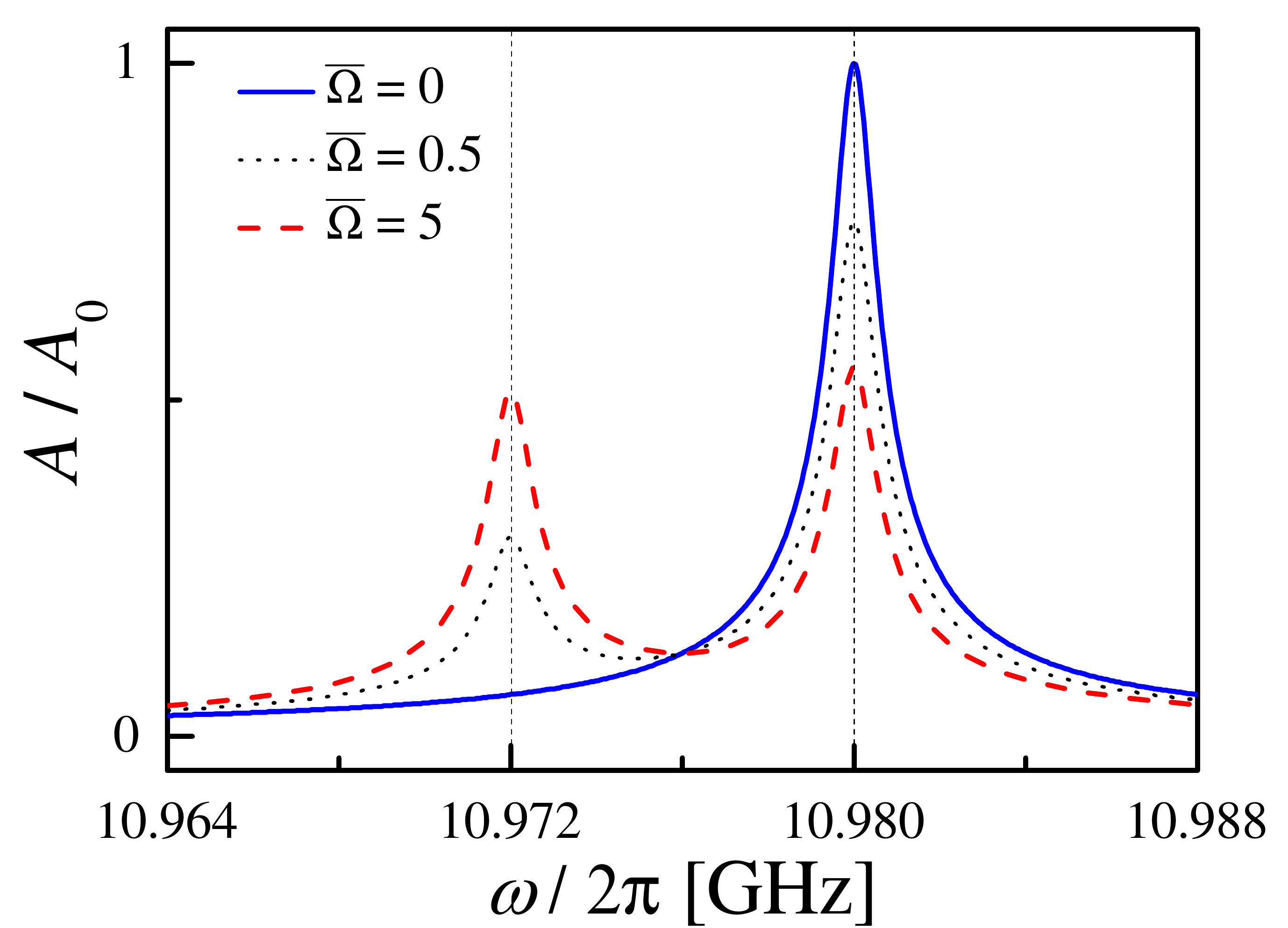}
\caption{Transmission amplitude for monitoring the state of a driven qubit.
The frequency $\protect\omega $ is in the range from $\protect\omega _{%
\mathrm{r}}-3\left\vert \protect\chi \right\vert $ to $\protect\omega _{%
\mathrm{r}}+3\left\vert \protect\chi \right\vert $. The peak corresponding
to the ground state is at $\protect\omega _{\mathrm{r}}+\left\vert \protect%
\chi \right\vert =10.98$~GHz$\cdot 2\protect\pi $. The peak appearing for
non-zero occupation of the excited state is at $\protect\omega =\protect%
\omega _{\mathrm{r}}-\left\vert \protect\chi \right\vert =10.972$~GHz$\cdot 2%
\protect\pi $. The height of the latter is defined by the normalized driving
amplitude $\overline{\Omega }$.}
\label{Fig:two-peaks}
\end{figure}

\begin{figure}[t]
\includegraphics[width=8cm]{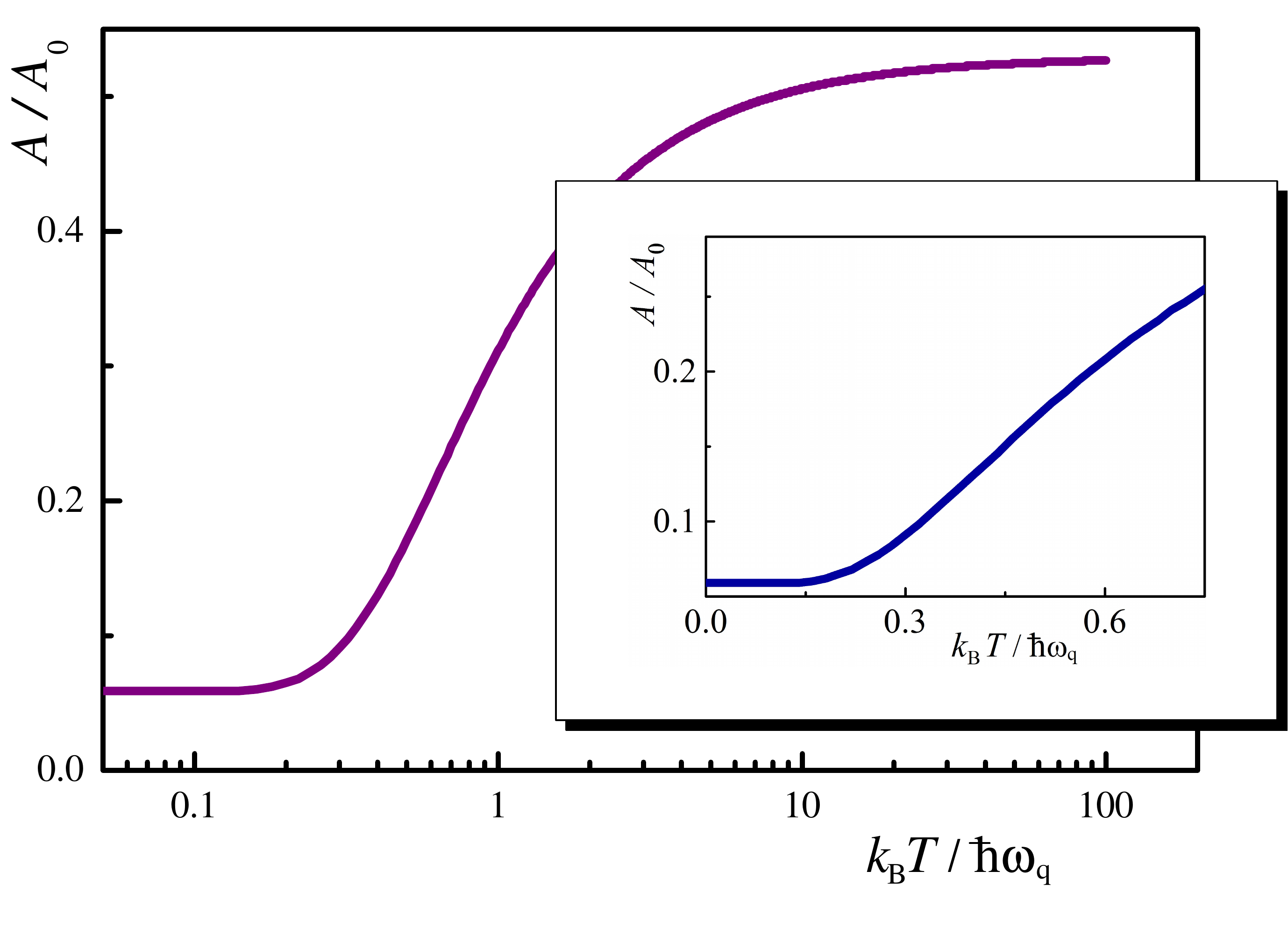}
\caption{Temperature dependence of the transmission amplitude~$A$ at the
frequency corresponding to the excited-state peak, which is $\protect\omega =%
\protect\omega _{\mathrm{r}}-\left\vert \protect\chi \right\vert $ in the
previous figure. Inset demonstrates the low-temperature region.}
\label{Fig:thermometer}
\end{figure}

With these equations (\ref{resonant}) and~(\ref{T-excited}) we calculate the
transmission amplitude, when the qubit was either resonantly driven (Fig.~%
\ref{Fig:two-peaks}) or in a thermal equilibrium (Fig.~\ref{Fig:thermometer}%
), respectively. For the former case we plot the frequency dependence of the
transmission amplitude in Fig.~\ref{Fig:two-peaks}. Similar dependence would
be for varying temperature; in Fig.~\ref{Fig:thermometer} we rather present
the transmission amplitude versus temperature for a fixed frequency $\omega
=\omega _{\mathrm{r}}+\chi =\omega _{\mathrm{r}}-\left\vert \chi \right\vert
$, where the excited-state peak appears. For calculations we took here the
parameters close to the ones of Ref.~[\onlinecite{Jin15}]:~$\omega _{\mathrm{%
r}}/2\pi =10.976$~GHz, $\omega _{\mathrm{q}}/2\pi =4.97$~GHz, $\chi /2\pi
=-4 $~MHz, and we have chosen$\ \varkappa /2\pi =1$~MHz. Again, as above, we
observe strong dependence on temperature. Advantages of probing qubit state
in a similar manner were discussed in Ref.~[\onlinecite{Reed10}]. There, it
was proposed to probe a \textit{driven} qubit state, while our proposal here
relates to the \textit{thermal}-equilibrium measurement and consists in
providing sensitive tool for thermometry. Indeed, similar temperature
dependence was recently observed by Jin et al. in Ref.~[\onlinecite{Jin15}].
In that work the authors studied the excited-state occupation probability in
a transmon with variable temperature. For comparison with that publication,
in the inset of Fig.~\ref{Fig:thermometer} we also present the
low-temperature region, with linear scale.

\section{Memcapacitance}

Now, having reached the agreement of the theory with the experiments, we
wish to explore other applications. In this section we mean possibilities
for memory devices, such as memcapacitors.

In general, a memory device with the input $u(t)$ and the output $y(t)$, by
definition, is described by the following relations \cite{Diventra09}
\begin{eqnarray}
y(t) &=&g(\mathbf{x},u,t)u(t),  \label{y(t)} \\
\mathbf{\dot{x}} &=&\mathbf{f}(\mathbf{x},u,t).  \label{x_dot}
\end{eqnarray}%
Here $g$ is the response function, while the vector function $\mathbf{f}$
defines the evolution of the internal variables, denoted as a vector $%
\mathbf{x}$. Depending on the choice of the circuit variables $u$ and $y$,
the relations (\ref{y(t)}-\ref{x_dot}) describe memristive, meminductive, or
memcapacitive systems. Relevant for our consideration is the particular case
of the voltage-controlled memcapacitive systems, defined by the relations
\begin{eqnarray}
q(t) &=&C_{\mathrm{M}}(\mathbf{x},V,t)V(t),  \label{q(t)} \\
\mathbf{\dot{x}} &=&\mathbf{f}(\mathbf{x},V,t).  \label{x_dot_2}
\end{eqnarray}%
Here the response function $C_{\mathrm{M}}$ is called the memcapacitance.

Relations (\ref{y(t)}-\ref{x_dot}) and their particular case, Eqs.~(\ref%
{q(t)}-\ref{x_dot_2}), were related to diverse systems, as described e.g. in
the review article~[\onlinecite{Pershin11}]. It was shown that the
reinterpretation of known phenomena in terms of these relations makes them
useful for memory devices. However, until recently their quantum analogues
remained unexplored. Then, some similarities and distinctions from classical
systems were analyzed in Refs.~[%
\onlinecite{Pfeiffer16, Shevchenko16, Salmilehto16,
Li16, Sanz17}]. In particular, it was argued that in the case of quantum
systems, the circuit input and output variables $u$ and $y$ should be
interpreted as quantum-mechanically averaged values or in the ensemble
interpretation \cite{Shevchenko16, Salmilehto16}. Detailed analysis of
diverse systems \cite{Pfeiffer16, Shevchenko16, Salmilehto16, Li16, Sanz17}
demonstrated that, being described by relations (\ref{y(t)}-\ref{x_dot}),
quantum systems could be considered as quantum memristors, meminductors, and
memcapacitors. These indeed displayed the pinched-hysteresis loops for
periodic input, while the frequency dependence may significantly differ from
the related classical devices. The former distinction is due to the
probabilistic character of measurements in quantum mechanics. Note that the
\textquotedblleft pinched-hysteretic loop\textquotedblright\ dependence is
arguably the most important property of memristors, meminductors, and
memcapacitors.\cite{Diventra09, Pershin11}

It is thus our goal in this section to demonstrate how the evolution
equations for a qubit-resonator system can be written in the form of the
memcapacitor relations, Eqs.~(\ref{q(t)}-\ref{x_dot_2}). This would allow us
to identify the related input and output variables, the internal-state
variables, the response and evolution functions. As a further evidence, we
will demonstrate one particular example, when for a resonant driving the
pinched-hysteresis loop appears.

The transmon treated as a memcapacitor is depicted in Fig.~\ref%
{Fig:memcapacitor}(a). As an input of such a memcapacitor we assume the
resonator antinode voltage $V$ (how a transmon is coupled to a
transmission-line resonator was shown in Fig.~\ref{Fig:vs_time_1}), while
the output is the charge $q$ on the external plate of the gate capacitor $C_{%
\mathrm{g}}$. One should differentiate between the externally applied
voltage, $V_{\mathrm{g}}=V_{A}\sin \omega t$, and the quantized antinode
voltage,
\begin{equation}
V=\left\langle \widehat{V}\right\rangle =V_{\mathrm{rms}}\left\langle
ae^{-i\omega t}+a^{\dag }e^{i\omega t}\right\rangle =2V_{\mathrm{rms}}\text{%
Re}\left\langle ae^{-i\omega t}\right\rangle ,  \label{V}
\end{equation}%
where $V_{\mathrm{rms}}=\sqrt{\hbar \omega _{\mathrm{r}}/2C_{\mathrm{r}}}$
is the root-mean-square voltage of the resonator, defined by its resonant
frequency $\omega _{\mathrm{r}}$ \ and capacitance $C_{\mathrm{r}}$.\cite%
{Koch07} This makes the difference from a charge qubit coupled directly to a
gate, such as in Ref.~[\onlinecite{Shevchenko16}]. Accordingly to Eq.~(\ref%
{V}), the voltage is related to the measurable values, the resonator output
field quadratures, Eq.~(\ref{IQ}). The charge $q$ is related to the voltage $%
V$ and the island charge $2e\left\langle n\right\rangle $ ($\left\langle
n\right\rangle $ is the average Cooper-pair number) as following \cite%
{Shevchenko16}:

\begin{equation}
q=C_{\mathrm{geom}}V+\frac{C_{\mathrm{g}}}{C_{\Sigma }}2e\left\langle
n\right\rangle \equiv C_{\mathrm{M}}V,  \label{q}
\end{equation}%
where we formally introduced the memcapacitance $C_{\mathrm{M}}$ as a
proportionality coefficient between the input $V$ and the output $q$. Given
the leading role of the shunt capacitance, here we have $C_{\Sigma
}=C_{J}+C_{\mathrm{g}}+C_{B}\sim C_{B}$ and $C_{\mathrm{geom}}=C_{\mathrm{g}%
}(C_{J}+C_{B})/C_{\Sigma }\approx C_{\mathrm{g}}$. The number operator $n$
is defined by the qubit Pauli matrix $\sigma _{y}$: $n=\frac{1}{4}\sqrt{%
\hbar \omega _{_{\mathrm{q}}}/E_{\mathrm{c}}}\sigma _{y}$. This allows us
rewriting Eq.~(\ref{q}),

\begin{equation}
\widetilde{q}=\text{Re\negthinspace }\left\langle a\right\rangle \cos \omega
t-\text{Im\negthinspace }\left\langle a\right\rangle \sin \omega t+\lambda
\left\langle \sigma _{y}\right\rangle ,  \label{q_}
\end{equation}%
where $\widetilde{q}=q/2C_{\mathrm{g}}V_{\mathrm{rms}}$ and $\lambda
=(e/4C_{\Sigma }V_{\mathrm{rms}})\sqrt{\hbar \omega _{_{\mathrm{q}}}/E_{%
\mathrm{c}}}$. We note that in related experiments, not only the quadratures
$I$ and $Q$ (which define Re$\left\langle a\right\rangle $ and Im$%
\left\langle a\right\rangle $), but also the qubit state, defined by the
values $\left\langle \sigma _{z}\right\rangle $ and $\left\langle \sigma
_{y}\right\rangle $, can be reliably probed, see Refs. [%
\onlinecite{Filipp09, Bianchetti09, McClure16, Gong16, Jerger17}].
Importantly, the memcapacitor's dynamics, i.e.~$q(t)$, is defined by rich
dynamics of both the resonator and the qubit, via $\left\langle
a\right\rangle $ and $\left\langle \sigma _{y}\right\rangle $, respectively.

\begin{figure}[t]
\includegraphics[width=8cm]{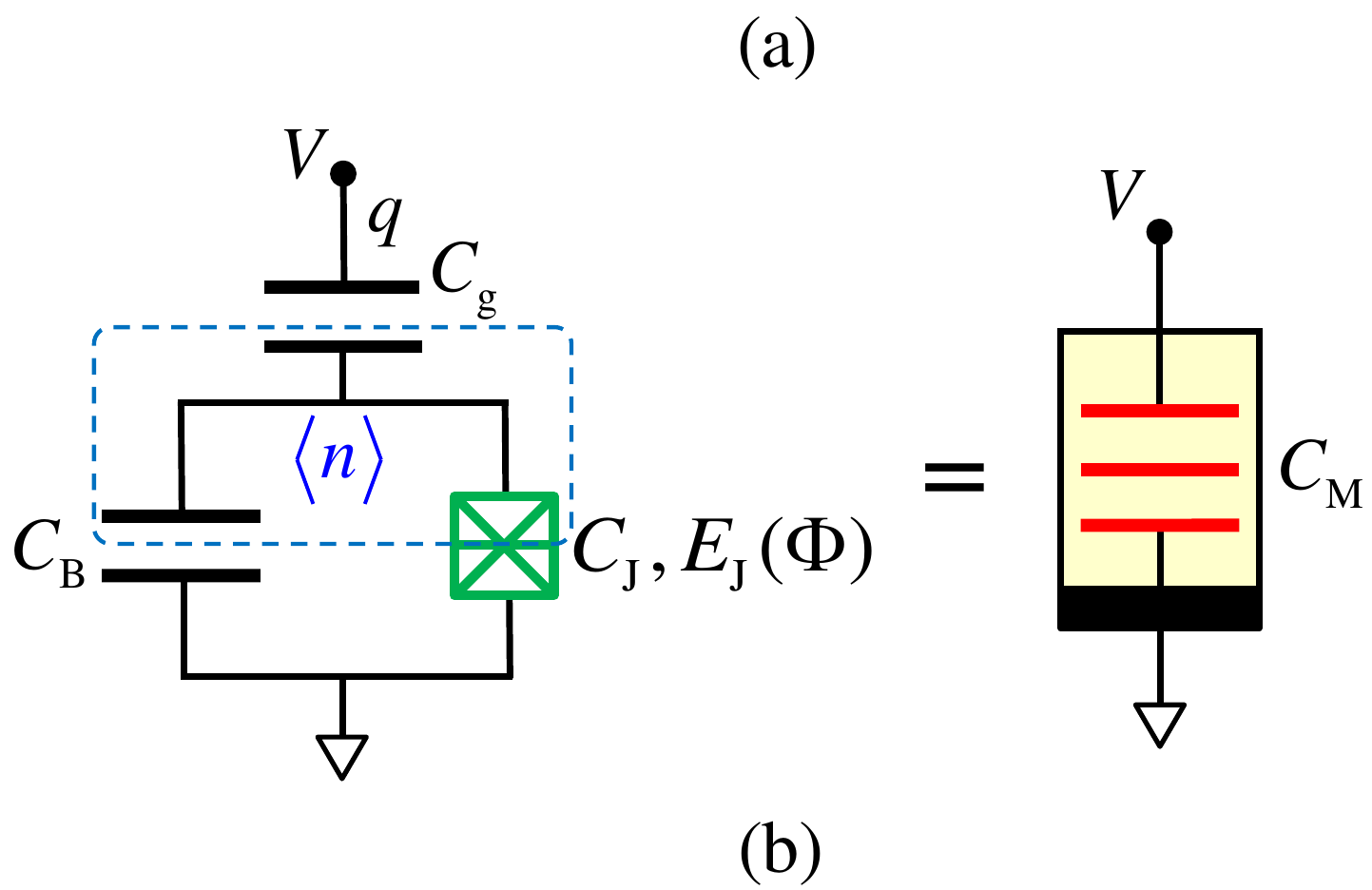} %
\includegraphics[width=8cm]{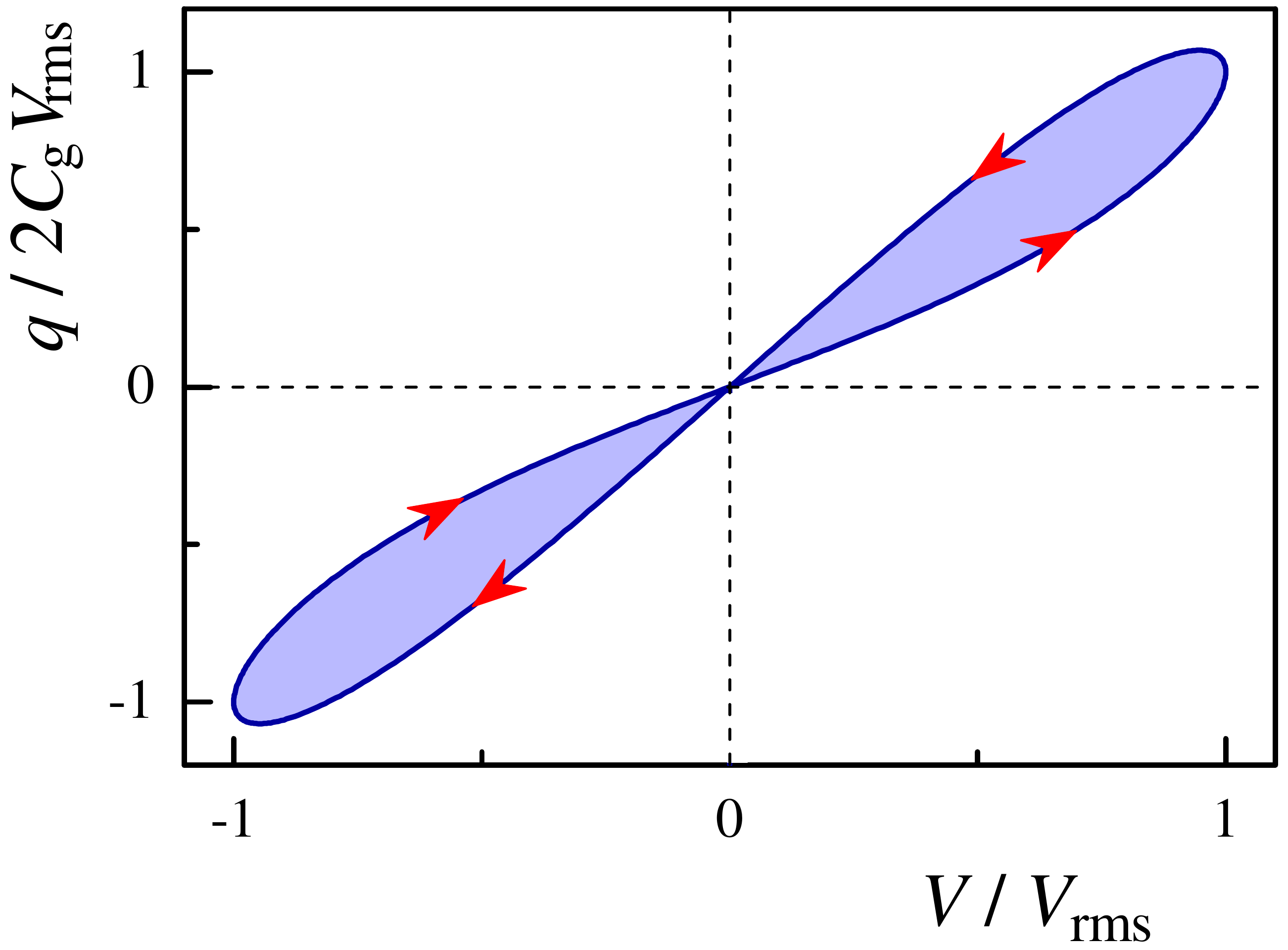}
\caption{A transmon qubit can be accounted as a memcapacitor. (a) Scheme of
a transmon-type qubit is shown to be equivalent to a memcapacitor $C_{%
\mathrm{M}}$, of which the symbol is shown to the right. (b)
Pinched-hysteretic curve in the voltage-charge plane as a fingerprint of a
memcapacitive behaviour.}
\label{Fig:memcapacitor}
\end{figure}

Importantly, here we have written the transmon-resonator equations in the
form of the memcapacitor first relation, Eq.~(\ref{q(t)}). We can see that
the role of the internal variables is played by the qubit charge $%
\left\langle n\right\rangle $. In its turn, the qubit state is defined by
the Lindblad equation, which now takes place of the second memcapacitor
relation, Eq.~(\ref{x_dot_2}). Such formulation demonstrates that our
qubit-resonator system can be interpreted as a quantum memcapacitor, which
is schematically displayed in Fig.~\ref{Fig:memcapacitor}(a).

The above formulas allow us to plot the charge--versus--voltage diagram. For
this we assume now that the qubit is driven by the field with amplitude $%
\Omega $ and frequency $\omega _{\mathrm{d}}$, which is resonant, $\omega _{%
\mathrm{d}}=\omega _{\mathrm{q}}(\Phi )$. This induces Rabi oscillations in
the qubit with the frequency $\Omega $. By numerically solving Eqs.~(5a-5d)
in Ref.~[\onlinecite{Bianchetti09}], we plot the $q$ versus $V$ diagram in
Fig.~\ref{Fig:memcapacitor}(b). To obtain the pinched-hysteresis-type loop,
we take the driving amplitude, $\Omega =2\omega _{\mathrm{r}}$, which
corresponds to the strong-driving regime. Other system parameters are the
same as used above (the ones of Ref.~[\onlinecite{Bianchetti09}]) and $%
\lambda =0.2$. In addition we have taken Re$\left\langle a\right\rangle =0$
and Im$\left\langle a\right\rangle =1$. Here we note that \negthinspace
\negthinspace $\left\langle a\right\rangle $ is a slow function of time, as
demonstrated in Fig.~\ref{Fig:vs_time_1}. (The characteristic time for this
is $2\pi /\varkappa $, which is $\gg $ $2\pi /\Omega $.) Moreover, the
diagram is not only defined by $\lambda $ (which is a constant), but also by
$\left\langle a\right\rangle $ (which can be adjusted, for example, by
choosing a moment of time in Fig.~\ref{Fig:vs_time_1}); so, for a different
value of $\lambda $ another value of $\left\langle a\right\rangle $ can be
taken. We note that the shaded area in Fig.~\ref{Fig:memcapacitor}(b) equals
to the energy consumed by the memcapacitor, $\int VIdt$.\cite{Pershin11}

Note that we made use of the two-level approximation for the transmon, and,
on the other hand considered the strong-driving regime, where $\Omega
=2\omega _{\mathrm{r}}$. This was needed for demonstrating the
pinched-hysteresis loop by illustrative means. While the strong-driving
regime was demonstrated in many types of qubits, in the transmon ones this
is complicated due to the weak anharmonicity, which may result in
transitions to the upper levels (cf. Refs.~[\onlinecite{Gong16, Lu17, Dai17}%
] though). In this way, one would have to confirm the calculations with the
more elaborated ones, by taking into account the higher levels (as e.g. in
Refs.~[\onlinecite{Peterer15, Pietikainen17a, Pietikainen17b}], see also our
discussion below, in Appendix~C) and clarify the relation, needed for the
hysteretic-type loops. Alternatively, one may think of the readily observed
Rabi oscillations in the megahertz domain and combine these with the
oscillations related to another resonator. At such low frequency the
resonator may be considered as classical, similarly to calculations in Ref.~[%
\onlinecite{Shevchenko16}].

\section{Conclusion}

We have considered the qubit-resonator system, accentuating on the situation
with a transmon-type qubit in a transmission-line resonator. The most
straightforward approach is the semiclassical theory, when all the
correlators are assumed to factorize, which has the advantage of getting
transparent analytical equations and formulas. We demonstrated that with
this we can describe relevant experiments~\cite{Bianchetti09, Jin15,
Pietikainen17a}. On the other hand, the validity of the semiclassical theory
was checked with the approach taking into account the two-operator
qubit-photon correlators, so-called semi-quantum approach. Furthermore, we
included temperature into consideration and studied its impact on the
measurable quadratures of the transmitted field. Due to the qubit-resonator
entanglement, the resonator transmission bears information about the
temperature experienced by the qubit. Consideration of this application, the
thermometry, was followed by another one, the memory device, known as a
memcapacitor. As a proof-of-concept, we demonstrated the pinched hysteretic
loop in the charge-voltage plane, the fingerprint of memcapacitance. In the
case with qubits, this loop is related to the Rabi-type oscillations. We
believe that such quantum memcapacitors, along with quantum meminductors and
memristors, will add new functionality to the toolbox of their classical
counterparts.

\begin{acknowledgments}
We are grateful to A.~Fedorov for stimulating discussions and critical
comments, to S.~Ashhab for critically reading the manuscript and for the
comments, and to Y.~V.~Pershin and E.~Il'ichev for fruitful discussions.
S.N.S. acknowledges the hospitality of School of Mathematics and Physics of
the University of Queensland, where part of this work was done; D.S.K.
acknowledges the hospitality of Leibnitz Institute of Photonic Technology.
This work was partly supported by the State Fund for Fundamental Research of
Ukraine (project \#~F66/95-2016) and DAAD bi-nationally supervised doctoral
degree program (grant \#~57299293).
\end{acknowledgments}

\appendix

\section{Lindblad and Maxwell-Bloch equations}

Consider how starting from the Hamiltonian (\ref{JC}), we get the motion
equations in the semiclassical approximation and obtain the steady-state
value for the photon operator in Eq.~(\ref{a}).

First, the Hamiltonian (\ref{JC}) is transformed with the operator $U=\exp %
\left[ i\omega t\left( a^{\dag }a+\sigma _{z}/2\right) \right] $ to the
following $H^{\prime }=UHU^{\dag }+i\hbar \dot{U}U^{\dag }$ (see e.g.~Ref.~[%
\onlinecite{Shevchenko14}]):%
\begin{equation}
H^{\prime }=\hbar \delta \omega _{\mathrm{r}}a^{\dag }a+\hbar \frac{\delta
\omega _{\mathrm{q}}}{2}\sigma _{z}+\hbar \mathrm{g}\left( \sigma a^{\dag
}+\sigma ^{\dag }a\right) +\hbar \xi \left( a^{\dag }+a\right) ,
\label{prime}
\end{equation}%
where%
\begin{equation}
\delta \omega _{\mathrm{r}}=\omega _{\mathrm{r}}-\omega ,~~\delta \omega _{%
\mathrm{q}}=\omega _{\mathrm{q}}-\omega .
\end{equation}

Then, the system's dynamics is described by the Lindblad master equation%
\begin{equation}
\dot{\rho}=-\frac{i}{\hbar }\left[ H^{\prime },\rho \right] +\varkappa
\mathcal{D}\left[ a\right] \rho +\Gamma _{1}\mathcal{D}\left[ \sigma \right]
\rho +\frac{\Gamma _{\phi }}{2}\mathcal{D}\left[ \sigma _{z}\right] \rho ,
\label{Lindblad}
\end{equation}%
where the damping terms model the loss of cavity photons at rate $\varkappa $%
, as well as the intrinsic qubit relaxation and pure dephasing at rates $%
\Gamma _{1}$ and $\Gamma _{\phi }$. The respective Lindblad damping
superoperators at non-zero temperature $T$ are given by \cite{ScullyZubairy}
\begin{subequations}
\label{D}
\begin{eqnarray}
\!\!\!\!\!\!\!\!\mathcal{D}\left[ a\right] \rho \!\! &=&\!\!\left( N_{%
\mathrm{th}}+1\right) \left( a\rho a^{\dag }-\frac{1}{2}\left\{ a^{\dag
}a,\rho \right\} \right) + \\
&&+N_{\mathrm{th}}\left( a^{\dag }\rho a-\frac{1}{2}\left\{ aa^{\dag },\rho
\right\} \right) ,  \notag \\
\!\!\!\!\!\!\!\!\mathcal{D}\left[ \sigma \right] \rho \!\! &=&\!\!\left( n_{%
\mathrm{th}}+1\right) \left( \sigma \rho \sigma ^{\dag }-\frac{1}{2}\left\{
\sigma ^{\dag }\sigma ,\rho \right\} \right) + \\
&&+n_{\mathrm{th}}\left( \sigma ^{\dag }\rho \sigma -\frac{1}{2}\left\{
\sigma \sigma ^{\dag },\rho \right\} \right) ,  \notag \\
\!\!\!\!\!\!\!\!\mathcal{D}\left[ \sigma _{z}\right] \rho \!\! &=&\!\!\left(
2n_{\mathrm{th}}+1\right) \left( \sigma _{z}\rho \sigma _{z}-\rho \right)
,\!\!\!
\end{eqnarray}%
\end{subequations}
\begin{equation}
N_{\mathrm{th}}\!\!=\!\!\left( \exp \!\!\left( \frac{\hbar \omega _{\mathrm{r%
}}}{k_{\mathrm{B}}T}\right) -1\right) ^{-1}\!\!,~n_{\mathrm{th}%
}\!\!=\!\!\left( \exp \!\!\left( \frac{\hbar \omega _{_{\mathrm{q}}}}{k_{%
\mathrm{B}}T}\right) -1\right) ^{-1}\!.
\end{equation}%
In particular, at $T=0$: $N_{\mathrm{th}}=n_{\mathrm{th}}=0$.

From the Lindblad equation~(\ref{Lindblad}), for the expectation values of
the operators $a$, $\sigma $, and $\sigma _{z}$ we obtain the following
system of equations (as in Refs.~[\onlinecite{Shevchenko14, Hauss08}]):
\begin{subequations}
\label{123}
\begin{eqnarray}
\!\!\!\!\!\!\!\!\!\!\!\!\frac{d\left\langle a\right\rangle }{dt}\!\!
&=&\!\!-i\delta \omega _{\mathrm{r}}^{\prime }\left\langle a\right\rangle -i%
\mathrm{g}\left\langle \sigma \right\rangle -i\xi ,  \label{i} \\
\!\!\!\!\!\!\!\!\!\!\!\!\frac{d\left\langle \sigma \right\rangle }{dt}\!\!
&=&\!\!-i\delta \omega _{\mathrm{q}}^{\prime }\left\langle \sigma
\right\rangle +i\mathrm{g}\left\langle a\sigma _{z}\right\rangle ,
\label{ii} \\
\!\!\!\!\!\!\!\!\!\!\!\!\frac{d\left\langle \sigma _{z}\right\rangle }{dt}%
\!\! &=&\!\!-i2\mathrm{g}\left( \left\langle a\sigma ^{\dag }\right\rangle
\!-\!\left\langle a^{\dag }\sigma \right\rangle \right) -\Gamma _{1}\left(
\!1\!+\!\frac{\left\langle \sigma _{z}\right\rangle }{z_{0}}\!\right) \!,
\label{iii}
\end{eqnarray}%
where
\end{subequations}
\begin{eqnarray}
\delta \omega _{\mathrm{r}}^{\prime } &=&\delta \omega _{\mathrm{r}}-i\frac{%
\varkappa }{2},\text{ }\delta \omega _{\mathrm{q}}^{\prime }=\delta \omega _{%
\mathrm{q}}-i\frac{\Gamma _{2}}{z_{0}},\text{ } \\
z_{0} &=&\tanh \left( \frac{\hbar \omega _{_{\mathrm{q}}}}{2k_{\mathrm{B}}T}%
\right) ,~\Gamma _{2}=\Gamma _{\phi }+\frac{\Gamma _{1}}{2}.  \notag
\end{eqnarray}%
The meaning of the value $z_{0}$ is in describing the qubit
temperature-dependent equilibrium population, which is seen from Eq.~(\ref%
{iii}), if neglecting the coupling $\mathrm{g}$.

The system of equations~(\ref{123}) becomes closed under the assumption that
all the correlation functions factorize (e.g. Ref.~[\onlinecite{Hauss08}]).
Then for the classical variables
\begin{equation}
\alpha =\left\langle a\right\rangle ,\quad s=\left\langle \sigma
\right\rangle ,\quad s_{z}=\left\langle \sigma _{z}\right\rangle
\end{equation}%
we obtain the equations, which are also called the Maxwell-Bloch equations,
\begin{subequations}
\label{Maxwell-Bloch}
\begin{eqnarray}
\dot{\alpha} &=&-i\delta \omega _{\mathrm{r}}^{\prime }\alpha -i\mathrm{g}%
s-i\xi , \\
\dot{s} &=&-i\delta \omega _{\mathrm{q}}^{\prime }s+i\mathrm{g}\alpha s_{z},
\\
\dot{s}_{z} &=&-i2\mathrm{g}\left( \alpha s^{\ast }-\alpha ^{\ast }s\right)
-\Gamma _{1}\left( 1+\frac{s_{z}}{z_{0}}\right) .
\end{eqnarray}%
This system of equations is convenient for describing the dynamics, as we do
in the main text. Also, these equations are simplified for the steady state,
where the time derivatives in the l.h.s. are zeros. Then for $\alpha $ and $%
s_{z}$, we obtain
\end{subequations}
\begin{eqnarray}
\alpha &=&-\xi \frac{\delta \omega _{\mathrm{q}}^{\prime }}{s_{z}\mathrm{g}%
^{2}+\delta \omega _{\mathrm{q}}^{\prime }\delta \omega _{\mathrm{r}%
}^{\prime }}, \\
s_{z} &=&-z_{0}+\frac{2z_{0}}{\Gamma _{1}}\left( \varkappa \left\vert \alpha
\right\vert ^{2}+2\xi \text{Im}\alpha \right) .
\end{eqnarray}%
These are further simplified in the low probing-amplitude limit. In this
case we note that $\alpha \sim \xi $ and obtain%
\begin{eqnarray}
s_{z} &=&-z_{0}, \\
\alpha &=&\xi \frac{\delta \omega _{\mathrm{q}}^{\prime }}{z_{0}\mathrm{g}%
^{2}-\delta \omega _{\mathrm{q}}^{\prime }\delta \omega _{\mathrm{r}%
}^{\prime }}.
\end{eqnarray}%
These formulas are analyzed in the main text.

\section{Semi-quantum model with temperature}

Here, following Refs.~[\onlinecite{Bianchetti09, Andre09}], we obtain
equations in the so-called semi-quantum model. This model essentially takes
into account the two-operator correlations, which were ignored in the
semiclassical approximation above.

\begin{figure}[t]
\includegraphics[width=8cm]{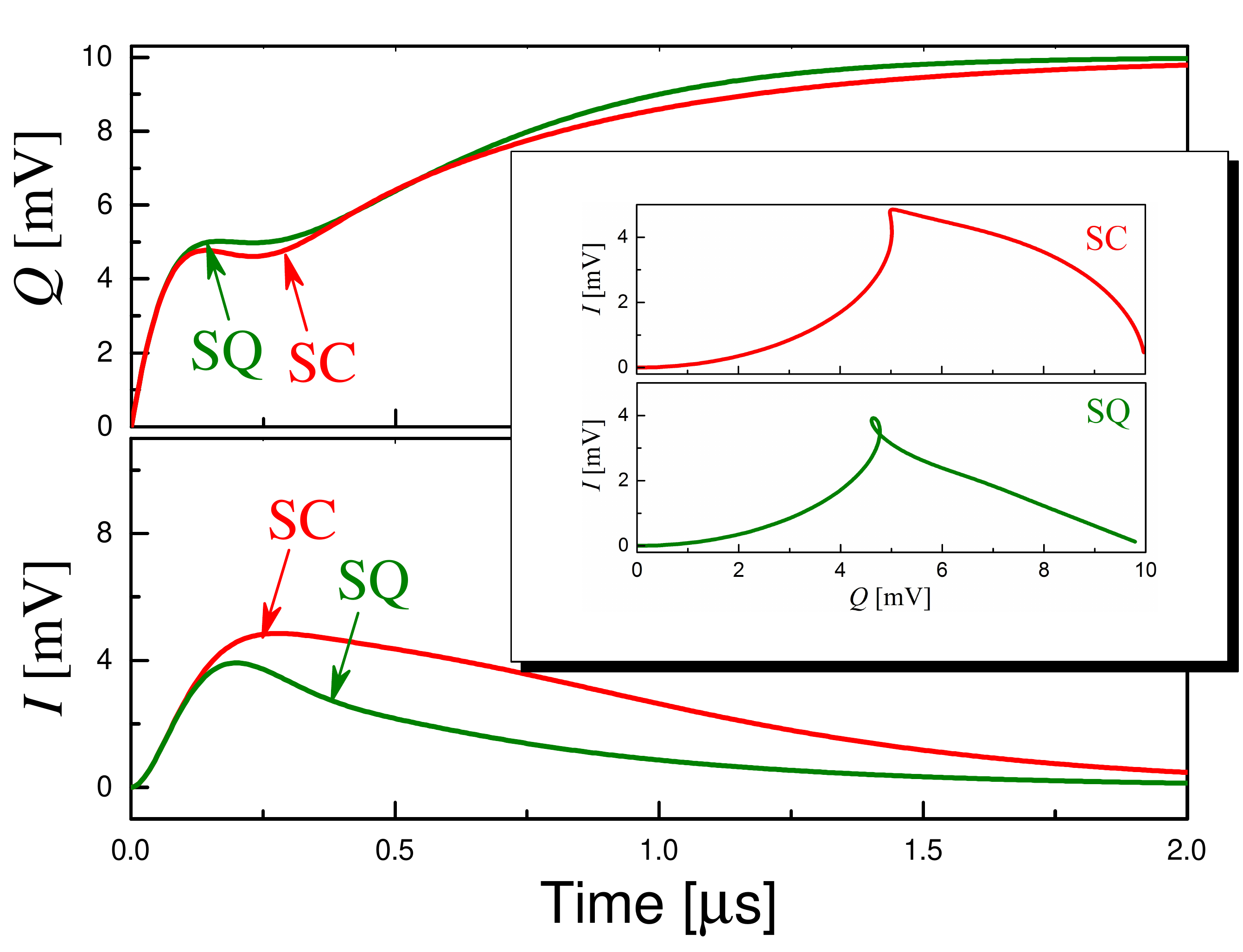}
\caption{Time dependence of the quadratures calculated in semiclassical (SC)
and semi-quantum (SQ) approximations, where upper panel presents $Q$
quadrature and lower panel presents $I$ quadrature. In the inset, dynamics
of the quadratures is presented in the $IQ$ plane, where the upper panel
presents the semiclassical approximation and the lower panel presents the
semi-quantum calculations.}
\label{Fig:IQ}
\end{figure}

We consider the situation, when the qubit-resonator detuning $\Delta =\hbar
\left( \omega _{\mathrm{q}}-\omega _{\mathrm{r}}\right) $ is much larger
than the coupling strength $\mathrm{g}$, then the system is described by the
dispersive approximation of the Jaynes-Cummings Hamiltonian \cite{Koch07,
Bianchetti09}

\begin{eqnarray}
H &=&\hbar (\omega _{\mathrm{r}}+\chi \sigma _{z})a^{\dag }a+\hbar \frac{%
\omega _{\mathrm{q}}+\chi }{2}\sigma _{z}+  \notag \\
&&+\hbar \left( \xi a^{\dagger }e^{-i\omega t}+\frac{\Omega }{2}\sigma
^{\dagger }e^{-i\omega _{\mathrm{d}}t}+h.c.\right) .
\end{eqnarray}
Here the second line represents the two control fields. The full Hamiltonian
$H$ of the system can be transformed with the operator $U=\exp [it(\omega
a^{\dagger }a+\omega _{\mathrm{d}}\sigma _{z}/2)]$ to the following $%
H^{\prime }=UHU^{\dagger }+i\hbar \dot{U}U^{\dagger }$

\begin{eqnarray}
H^{\prime } &=&\hbar (\delta \omega _{\mathrm{r}}+\chi \sigma _{z})a^{\dag
}a+\hbar \frac{\delta \omega _{\mathrm{q-d}}+\chi }{2}\sigma _{z}+  \notag \\
&&+\hbar \xi (a^{\dagger }+a)+\hbar \frac{\Omega }{2}(\sigma +\sigma
^{\dagger }),
\end{eqnarray}%
where $\delta \omega _{\mathrm{q-d}}=\omega _{\mathrm{q}}-\omega _{\mathrm{d}%
}$. Following Ref.~[\onlinecite{Bianchetti09}], now for the non-zero
temperature, from the Lindblad equation~(\ref{Lindblad}), for the
expectation values of the operators $\langle \sigma _{i}\rangle $ $(i=x,y,z)$
and the resonator field operators $\langle a\sigma _{i}\rangle $ and $%
\langle a^{\dagger }a\rangle $ we obtain the system of equations:

\begin{subequations}
\label{SQ}
\begin{eqnarray}
\frac{d}{dt}\langle \sigma _{z}\rangle \!\!\! &=&\!\!\!\Omega \langle \sigma
_{y}\rangle -\Gamma _{1}\left( 1+\frac{\langle \sigma _{z}\rangle }{z_{0}}%
\right) , \\
\frac{d}{dt}\langle \sigma _{x}\rangle \!\!\! &=&\!\!\!-\!\left( 2\chi
\!\!\left\langle a^{\dagger }a\right\rangle \!+\!\delta \omega _{\mathrm{q-d}%
}+\!\chi \right) \!\langle \sigma _{y}\rangle -\Gamma _{2}\!\frac{\langle
\sigma _{x}\rangle }{z_{0}}\!, \\
\frac{d}{dt}\langle \sigma _{y}\rangle \!\!\! &=&\!\!\!\left( 2\chi
\left\langle a^{\dagger }a\right\rangle +\delta \omega _{\mathrm{q-d}}+\chi
\right) \langle \sigma _{x}\rangle - \\
&&-\Gamma _{2}\frac{\langle \sigma _{y}\rangle }{z_{0}}-\Omega \langle
\sigma _{z}\rangle ,  \notag \\
\frac{d}{dt}\langle a\rangle \!\!\! &=&\!\!\!-i\left( \delta \omega _{%
\mathrm{r}}\langle a\rangle +\chi \langle a\sigma _{z}\rangle +\xi \right) -%
\frac{\kappa }{2}\langle a\rangle , \\
\frac{d}{dt}\langle a^{\dagger }a\rangle \!\!\! &=&\!\!\!-2\xi \text{%
\thinspace Im\thinspace }\langle a\rangle +\kappa \left( N_{\mathrm{th}%
}-\langle a^{\dagger }a\rangle \right) , \\
\frac{d}{dt}\langle a\sigma _{z}\rangle \!\!\! &=&\!\!\!-i\left( \delta
\omega _{\mathrm{r}}\left\langle a\sigma _{z}\right\rangle +\chi \langle
a\rangle +\xi \right) \langle \sigma _{z}\rangle + \\
&&+\Omega \langle a\sigma _{y}\rangle -\Gamma _{1}\left\langle
a\right\rangle -\left( \frac{\Gamma _{1}}{z_{0}}+\frac{\kappa }{2}\right)
\langle a\sigma _{z}\rangle ,  \notag \\
\frac{d}{dt}\langle a\sigma _{x}\rangle \!\!\! &=&\!\!\!-i\delta \omega _{%
\mathrm{r}}\left\langle a\sigma _{x}\right\rangle -\left( \frac{\Gamma _{2}}{%
z_{0}}+\frac{\kappa }{2}\right) \langle a\sigma _{x}\rangle + \\
&&+\left( \delta \omega _{\mathrm{q-d}}+2\chi \left( \langle a^{\dagger
}a\rangle +1\right) \right) \langle a\sigma _{y}\rangle -i\xi \langle \sigma
_{x}\rangle ,  \notag \\
\frac{d}{dt}\langle a\sigma _{y}\rangle \!\!\! &=&\!\!\!-i\delta \omega _{%
\mathrm{r}}\left\langle a\sigma _{y}\right\rangle -\left( \frac{\Gamma _{2}}{%
z_{0}}+\frac{\kappa }{2}\right) \langle a\sigma _{y}\rangle - \\
&&-i\xi \langle \sigma _{y}\rangle -\Omega \langle a\sigma _{z}\rangle -
\notag \\
&&-\left( \delta \omega _{\mathrm{q-d}}+2\chi \left( \langle a^{\dagger
}a\rangle +1\right) \right) \langle a\sigma _{x}\rangle .
\end{eqnarray}
\end{subequations}

Here we have truncated the infinite series of equations by factoring
higher-order terms $\langle a^{\dagger }a\sigma _{i}\rangle \thickapprox
\langle a^{\dagger }a\rangle \langle \sigma _{i}\rangle $ and $\langle
a^{\dagger }aa\sigma _{i}\rangle \thickapprox \langle a^{\dagger }a\rangle
\langle a\sigma _{i}\rangle $. Note that at $T=0$, the system (\ref{SQ})
coincides with Eq.~(5) in Ref.~[\onlinecite{Bianchetti09}].

We have numerically solved the system of equations~(\ref{SQ}) and the
results are shown in Fig.~\ref{Fig:IQ}. The two main panels present dynamics
of the quadratures, where the red curves are the result of calculations in
the semiclassical calculations, demonstrated in the main text in Fig.~\ref%
{Fig:vs_time_1}. The green curves present dynamics of the quadratures
calculated in the semi-quantum approximation. There is a good quantitative
agreement between the two approximations for the $Q$ quadrature, while the
agreement for the $I$ quadrature during the transient stage is only
qualitative. In addition, to further emphasize similarity and distinction of
the two approaches, we present these quadratures in the inset in Fig.~\ref%
{Fig:IQ}. While the two approaches give similar dynamics of the quadratures,
the semiclassical approximation does not describe the self-crossing of the $%
IQ$ curve. Such dependence, including the self-crossing feature, was
demonstrated in Fig.~4(d) of Ref.~[\onlinecite{Bianchetti09}]. We can make
the conclusion here that the semiclassical calculations are good for
obtaining analytical expressions, which describe qubit-resonator dynamics,
while for describing some fine features of the dynamics, semi-quantum
calculations may be necessary. Most importantly, we can see that the
semiclassical calculations give correct values for the stationary variables.

%Remarkably, our calculations gave the same results as the semiclassical model. In
%particular, we observed that changing the qubit temperature results in the
%resonance frequency shift, as in Fig.~\ref{Fig:phase_shift}, while changing
%only the effective temperature of the resonator results in the suppression
%of the resonant peaks, without their shift, as in Ref.~[\onlinecite{Fink10}].

\section{Quantum-to-classical transition for the strongly driven
qubit-resonator system}

In order to further demonstrate our approach, we devote this Section to the
regime of strong driving of the qubit-resonator system. The frequency shift
of the resonant transmission through the system was recently studied in
detail in Refs.~[\onlinecite{Pietikainen17a, Pietikainen17b}]. There, the
authors studied such quantum-to-classical transition both experimentally and
theoretically. Importantly, they have compared several numerical approaches,
with RWA and without, taking into account both two transmon levels only and
also higher levels. Our calculations are rather analytical and comparing
them with the ones from Refs.~[\onlinecite{Pietikainen17a,
Pietikainen17b}] shows both applicability and limitations of our approach.

\begin{figure}[t]
\includegraphics[width=8cm]{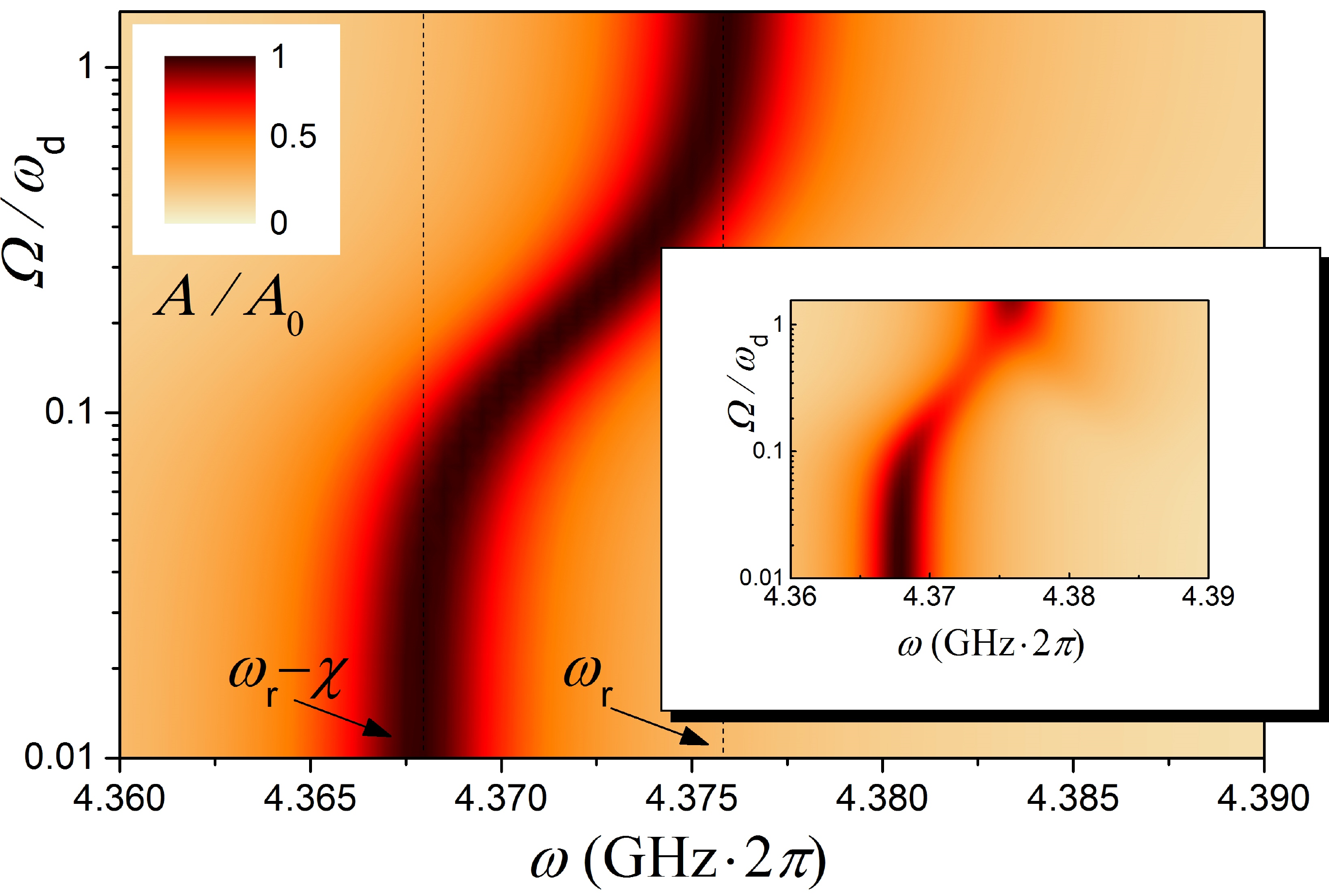}
\caption{The resonant-frequency shift for the strongly-driven
qubit-resonator system. The transmission amplitude $A$ is plotted as a
function of the probing frequency $\protect\omega $ and the driving
amplitude $\Omega $ for the off-resonant driving with $\protect\omega _{%
\mathrm{d}}\neq \protect\omega _{\mathrm{q}}$. Semiclassical and
semi-quantum calculations are presented in the main panel and in the inset,
respectively.}
\label{Fig:QtC}
\end{figure}

So, for calculations we took the parameters close to the ones of Ref.~[%
\onlinecite{Pietikainen17a}]:~$\omega _{\mathrm{r}}/2\pi =4.376$~GHz, $%
\omega _{\mathrm{q}}/2\pi =5.16$~GHz, $\mathrm{g}/2\pi =80~$MHz (which gives
$\chi /2\pi =8.2$~MHz), $\varkappa /2\pi =4$~MHz, $\Gamma _{1}/2\pi =2$~MHz,
$\Gamma _{2}=\Gamma _{1}$, and also driving frequency $\omega _{\mathrm{d}%
}=4.35$~GHz$\cdot 2\pi $. For this off-resonant driving ($\omega _{\mathrm{d}%
}\neq \omega _{\mathrm{q}}$) we make use of Eq.~(\ref{non-resonant}) and
then, together with Eq.~(\ref{general}), we plot the transmission amplitude
in the main panel in Fig.~\ref{Fig:QtC}. This displays transition from the
low-amplitude driving, when the resonant transmission appears around $\omega
=\omega _{\mathrm{r}}-\chi $, corresponding to the qubit in the ground
state, to the high-amplitude driving, when the qubit is in the superposition
state, with average $P_{+}=1/2$ and the resonant transition appears around $%
\omega =\omega _{\mathrm{r}}$. One can observe that with increasing the
driving amplitude $\Omega $, the frequency shifts by the value $\chi $,
which is defined in Eq.~(\ref{dispersive}), $\chi =\mathrm{g}_{0}^{2}E_{%
\mathrm{c}}/\Delta (\Delta -E_{\mathrm{c}})$. We must note that for the
resonant driving, with $\omega _{\mathrm{d}}=\omega _{\mathrm{q}}$, it is
much easier to saturate the qubit population and this happens at much
smaller driving power, at $\Omega \sim 0.001\omega _{\mathrm{d}}\sim 2\sqrt{%
\Gamma _{1}\Gamma _{2}}$, rather than at $\Omega \sim \omega _{\mathrm{d}}$
in Fig.~\ref{Fig:QtC}; it is thus non-resonant driving which allows
consideration of the resonance shift in the regime of strong driving~[%
\onlinecite{Jerger17}].

In addition to the semiclassical calculations, in the inset in Fig.~\ref%
{Fig:QtC} we present the resonant-frequency shift in the semi-quantum
approximation, for which we solved the system of equations~(\ref{SQ}).
Overall, the shift of the resonance is consistent with the semiclassical
calculations in the main part of Fig.~\ref{Fig:QtC}; the suppression of the
peak in the crossover region makes better resemblance with the experimental
results and numerical calculations in Ref.~[\onlinecite{Pietikainen17a}].

\nocite{apsrev41Control}
\bibliographystyle{apsrev4-1}
\bibliography{thermometry}

\end{document}